\documentclass[conference]{IEEEtran}
\IEEEoverridecommandlockouts
\usepackage{cite}

\usepackage{epsfig}
\usepackage{multirow}
\usepackage{microtype} 
\usepackage{booktabs}  
\usepackage{url}
\usepackage{paralist}
\usepackage{subfig}
\usepackage{amsmath}
\usepackage{wrapfig}
\usepackage{blkarray, bigstrut}
\usepackage{gensymb}
\usepackage{enumitem}
\usepackage[utf8x]{inputenc}

\usepackage[linesnumbered,ruled,vlined]{algorithm2e}

\usepackage{algorithmic}

\usepackage{color}

\usepackage[textsize=footnotesize]{todonotes}

\DeclareMathAlphabet{\mathitbf}{OML}{cmm}{b}{it}
\DeclareMathAlphabet{\mathbfit}{OML}{cmm}{b}{it}

\begin{document}
	

\title{Temperature-Based Hardware Trojan For Ring-Oscillator-Based TRNGs}

\author{
     \IEEEauthorblockN{Samaneh Ghandali\IEEEauthorrefmark{1}, Daniel Holcomb\IEEEauthorrefmark{1}, Christof Paar\IEEEauthorrefmark{1}\IEEEauthorrefmark{2}}
     \IEEEauthorblockA{\IEEEauthorrefmark{1}University of Massachusetts Amherst, USA
     \\\{samaneh, dholcomb\}@umass.edu}
     \IEEEauthorblockA{\IEEEauthorrefmark{2}Horst G\"ortz Institute for IT Security, Ruhr-Universit\"at Bochum, Bochum, Germany
     \\\{christof.paar\}@rub.de}
}


\maketitle



\begin{abstract}

True random number generators (TRNGs) are essential components of cryptographic designs, which are used to generate private keys for encryption and authentication, and are used in masking countermeasures. In this work, we present a mechanism to design a stealthy parametric hardware Trojan for a ring oscillator based TRNG architecture proposed by Yang et al. at ISSCC 2014. Once the Trojan is triggered the malicious TRNG generates predictable non-random outputs. Such a Trojan does not require any additional logic (even a single gate) and is purely based on subtle manipulations on the sub-transistor level. The underlying concept is to disable the entropy source at high temperature to trigger the Trojan, while ensuring that Trojan-infected TRNG works correctly under normal conditions. We show how an attack can be performed with the Trojan-infected TRNG design in which the attacker uses a stochastic Markov Chain model to predict its reduced-entropy outputs.  

\end{abstract}



\section{Introduction}
\label{sec:intro}

Cryptographic devices are those pieces of (usually) hardware that implement cryptographic algorithm(s) providing different aspects of security. Since such devices often deal with secret information and/or privacy of the users, they are very attractive target for subversion by malicious actors. Manipulating hardware implementations as opposed to software implementations can lead to cryptographic Trojans that are particularly difficult to detect. Hardware Trojans have gained high attention in academia and industry as well as government agencies, and can leak the secrets in a particular fashion without the notice of the end users.

Research concerning the Trojan design and Trojan detection are large and active.
Nevertheless, these two topics are closely related.
The effective detection mechanisms and countermeasures are only possible when there is an understanding of how hardware Trojans can be built. Amongst several different ways to insert a Trojan into an IC, we can refer to those conducted $i$) by an untrusted semiconductor foundry during manufacturing, $ii$) by the original hardware designer who is pressured by the government bodies, and $iii$) in the third-party IP cores.
Most of the hardware Trojans are inserted by modifying a few gates (can be done at different abstraction levels) \cite{DBLP:conf/ches/GhandaliBHP16}, \cite{EnderG0P17}.
In short, one of the main goals of the Trojans is to be designed/implemented in such a way that the chance of detection becomes very low.

High entropy random numbers are very essential component in many aspects of information security, which forms the foundation for many cryptographic algorithms. Some common applications are generating private keys, nonces, random numbers in challenge response protocols, and random numbers in side-channel leakage countermeasure implementations. One of the most popular methods for generating random numbers is sampling jittery signals generated by ring oscillators (ROs) \cite{yang201416} and \cite{DBLP:conf/ches/CherkaouiFFA13}. In this paper, we present a parametric hardware Trojan for an RO-based TRNG presented in \cite{yang201416} in such a way that it works correctly under normal environmental conditions, but it produces non-random and predictable outputs at particular environmental conditions such as high environmental temperature. Our Trojan does not require the addition of any additional logic (even a single gate) to the design, making it extremely hard to detect. More precisely, our technique injects a \textit{parametric} Trojan that can be triggered. Under normal conditions the randomness of the TRNG output is not affected, which enables the Trojan to avoid being detected by an evaluation lab. By increasing the temperature of the subverted device (or by increasing its workload) the Trojan is triggered and exhibits non-random and periodic outputs. We show that by injecting this Trojan, we are able to control the output of the TRNG. This biasing significantly lowers the security level even of highly protected crypto-core implementations rely on the TRNG. Also we elaborate a stochastic model based on Markov Chains by which the attacker’s knowledge enables predicting the output of the Trojan infected TRNG.

Section~\ref{sec:relatedwork} reviews related work in the areas of hardware Trojans.
Afterwards, in Section~\ref{sec:3edgetrng} we describe the ring-oscillator-based TRNG architecture which is our target to design the hardware Trojan. In Section~\ref{sec:Trojan} we express our core idea how to build and insert our Trojan into the ring-oscillator-based TRNG. In Section~\ref{sec:prediction} we explain how to elaborate a stochastic model based on Markov Chain for the attacker’s knowledge to predict the output of the Trojan infected TRNG. In Section~\ref{sec:result} the measurement results are provided. Finally, we conclude our work in Section~\ref{sec:conclude}.

\section{Related Work}
\label{sec:relatedwork}

Malicious and intentional modification of integrated circuit (IC) during manufacturing in untrusted foundry is an emerging security concern. This problem exists because the majority of ICs are fabricated abroad, and a government agency could force a foundry to manipulate the design maliciously. Also, an IC designer can be pressured by her own country government to modify the ICs maliciously, e.g., those ICs that are used in overseas products. Another possible insertion point are 3rd party IP cores. In general, a hardware Trojan is a back-door that can be inserted into an integrated circuit as an undesired and malicious modification, which makes  the behavior of the IC incorrect. 

There are many ways to categorize Trojans such as categorizing based on physical characteristics, design phase, abstraction level, location, triggering mechanism, and functionality. But a common Trojan categorization is based on the activation mechanism (Trojan trigger) and the effect on the circuit functionality (Trojan payload). A set of conditions that cause a Trojan to be activated is called trigger. Trojans can combinationally or sequentially be triggered. An attacker chooses a rare trigger condition so that the Trojan would not be triggered during conventional design-time verification and manufacturing test. Sequentially-triggered Trojans (time bombs) are activated by the occurrence of a rare sequence events, or after a period of continuous operation  \cite{ChakrabortyNB09}.

The goal of the Trojan can be achieved by payload which can change the circuit functionally or leak its secret information. In \cite{DBLP:conf/host/JinM08} a categorization method according to how the payload of a Trojan works has been defined; some Trojans after triggering, propagate internal signals to output ports which can reveal secret information to the attackers (explicit payload). Other Trojans may make the circuit malfunction or destroy the whole chip (implicit payload). Another categorization for actions of hardware Trojans has been presented in \cite{DBLP:conf/dft/WangSTP08}, in which the actions can be categorized into classes of modify functionality, modify specification, leak information, and denial of service.

The work in \cite{DBLP:conf/ches/BeckerRPB13} presented building stealthy Trojans at the layout-level. A hardware Trojan was inserted into a cryptographically-secure PRNG and into a side-channel resistant Sbox by manipulating the dopant polarity of a few registers. Building hardware Trojans that are triggered by aging was presented in \cite{DBLP:conf/ahs/ShiyanovskiiWRPWC10}. These Trojans only become active after the IC has been working for a long time. 

A class of hardware Trojans -- Malicious Off-chip Leakage Enabled by Side-channels (MOLES) -- has been presented in \cite{DBLP:conf/iccad/LinBP09}, which can retrieve secret information through side channels. They formulated the mechanism and detection methods of MOLES in theory and provided a verification process for multi-bit key extractions. A parametric Trojan has been introduced in \cite{DBLP:conf/fdtc/KumarJBP14} which triggers with a probability increasing under reduced supply voltage. In \cite{DBLP:conf/ches/GhandaliBHP16} a design methodology for building stealthy parametric hardware Trojans and its application to Bug Attacks \cite{DBLP:conf/crypto/BihamCS08} has been proposed. The Trojan is based on increasing delay of gates of a very rare-sensitized path in a combinatorial circuit, such as an arithmetic multiplier circuit. It is stealthy and has rare trigger conditions, so that the faulty behavior of the circuit under attack only occurs for very few combinations of the input vectors. Also an attack on the ECDH key agreement protocol by this Trojan has been presented in this work.

A physical attack on random number generators was presented in \cite{DBLP:conf/ches/MarkettosM09} which aims at an RO based TRNG implemented in an IC. Injecting a sine wave onto the power supply, the operating conditions were modified and a bias appeared at the output signal. Another physical attack presented in \cite{conf/cosade/BayonBAFPRM12}, targets another RO based TRNG \cite{DBLP:journals/ijrc/WoldT09} using an electromagnetic attack. In this attack, the ROs were locked on the injection frequency, generating a controllable bias at the output. The work in \cite{martin2015fault} investigated the impact of power and clock glitches, temperature and underpowering on a TRNG design \cite{DBLP:conf/ches/CherkaouiFFA13} implemented on an FPGA.

\section{Ring oscillator-based TRNG}
\label{sec:3edgetrng}

\begin{figure}[htbp]%
	\vspace{-0.1 in}
	\centering
	\includegraphics[width=1.0\columnwidth]{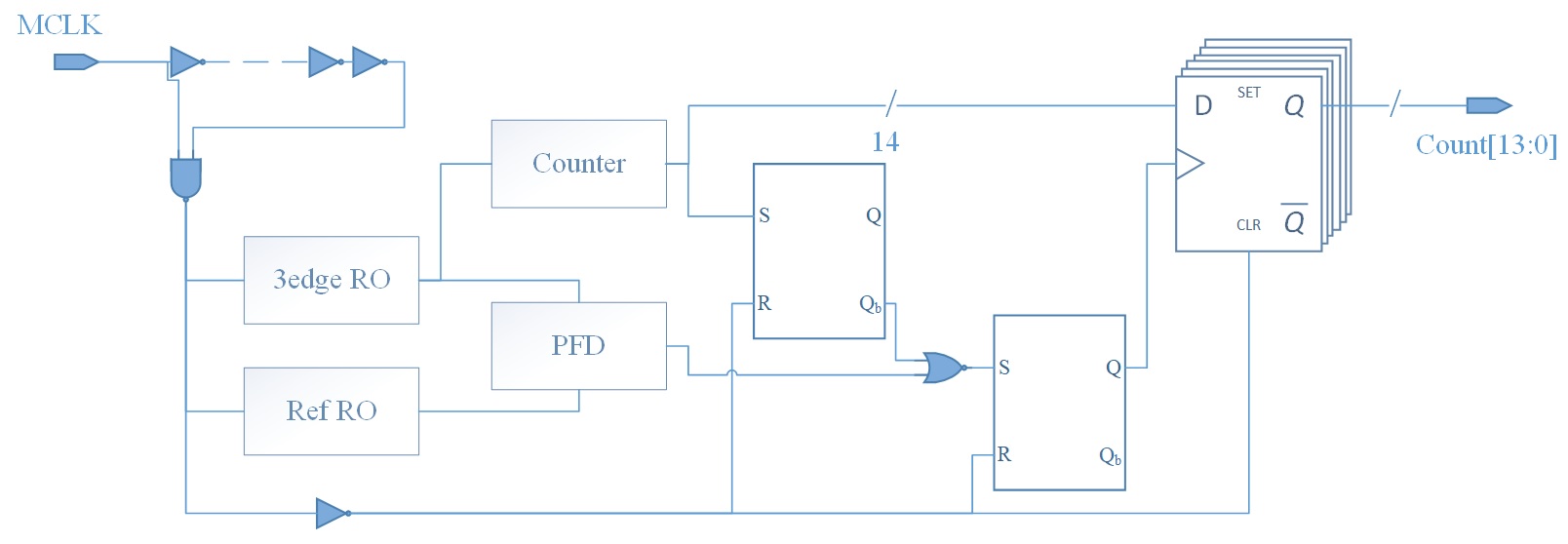}%
	\caption{TRNG system block diagram \cite{yang201416}}%
	\label{fig:TRNGarch}%
\end{figure}

\begin{figure}[htbp]%
	\centering
	\includegraphics[width=1.0\columnwidth]{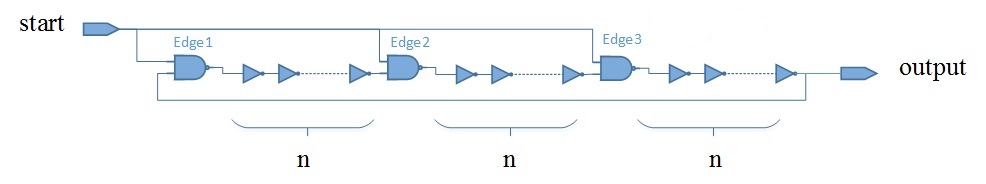}%
	\caption{3-edge ring oscillator}%
	\label{fig:3edgeRO}%
\end{figure}

\begin{figure}[htbp]%
	\centering
	\includegraphics[width=1.0\columnwidth]{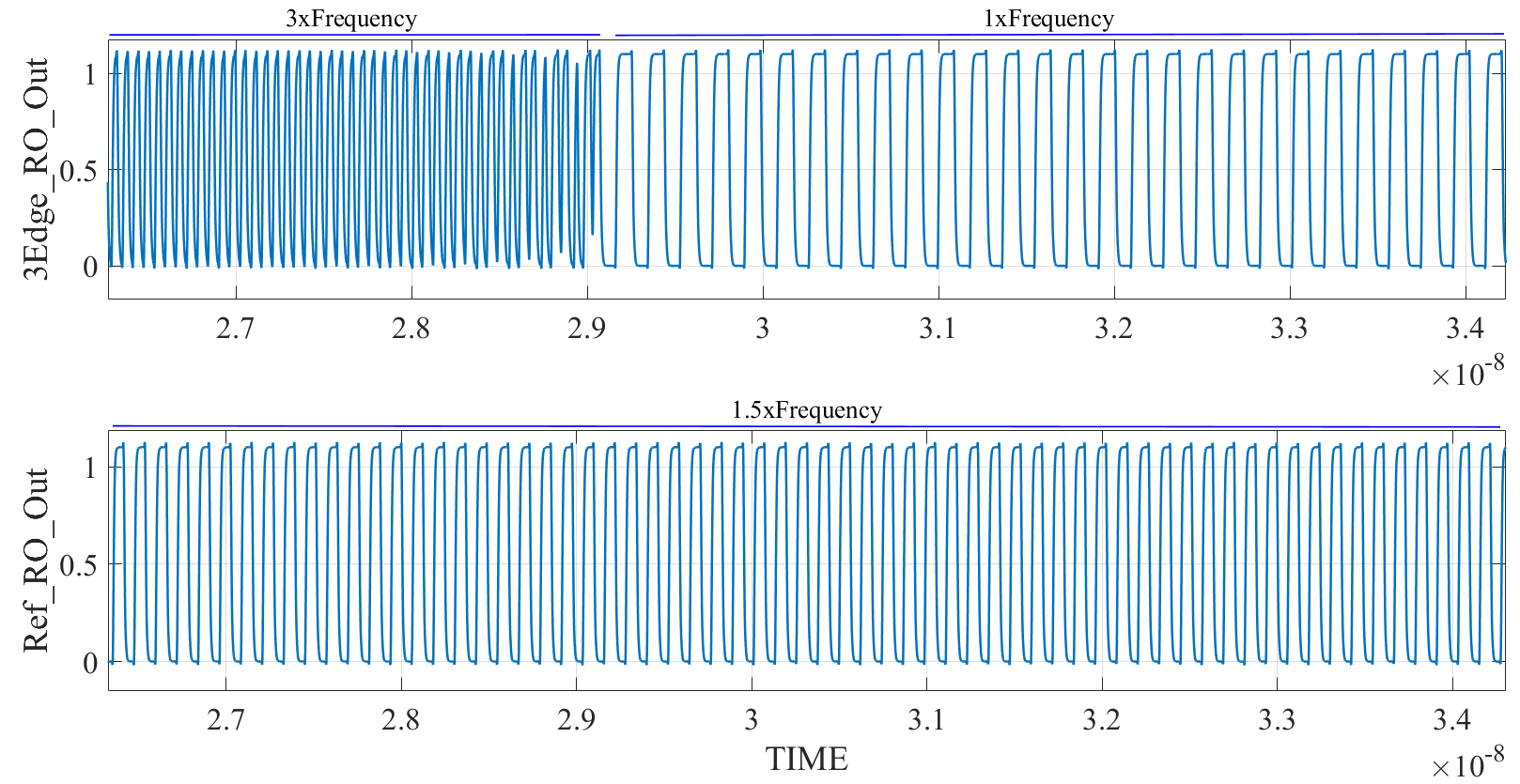}%
	\caption{Output waveforms of 3-edge RO (top) and regular RO (bottom)}%
	\label{fig:output_regular_3edge_RO}%
\end{figure}

We consider the true random number generator (TRNG) design proposed in \cite{yang201416}. Figure~\ref{fig:TRNGarch} shows the TRNG architecture, which is based on the collapse time of three racing edges in a ring oscillator (RO). The design has two ring oscillators (RO). The first one is a reference that operates as a standard single-edge ring oscillator. The second one, which is called 3-edge RO, has three edges injected by three input nodes that propagate through the ring together at the same time (Figure~\ref{fig:3edgeRO}). These edges in the 3-edge RO have same period, but they are shifted 120° in phase. As a result of this the frequency of the output of the 3-edge RO is boosted 3$ \times $ in comparison to the regular RO. There is an increasing variation of the pulse width between edges in the 3-edge RO because of thermal noise (jitter) that exists in the system. This variation in the pulse widths causes neighboring edges to eventually collapse in the 3-edge RO, after which there is only a single oscillation in the ring. The collapse event in turn causes the 3-edge RO to change to a typical 1x frequency mode as can be seen in Figure~\ref{fig:output_regular_3edge_RO}. The time to collapse is used as the entropy source for the TRNG.

Phase frequency detector (PFD) module in the TRNG architecture shown in Figure~\ref{fig:TRNGarch} is used to detect the edge collapse events by comparing the frequencies of the regular RO and the 3-edge RO. A 14-bit counter counts the number of cycles until edge collapse event. This counter increments on rising edges of the 3-edge RO. 


%

The number of cycles to collapse follows inverse Gaussian distribution caused by thermal noise. In this design effect of process variation is canceled because all three edges propagate through the same RO stages\cite{DBLP:conf/cicc/TangKLPK14}. We need to extract uniformly distributed random bits from collapse time. A simple method which has been applied in TRNG designs \cite{DBLP:conf/cicc/TangKLPK14}, \cite{liu2011true} is to take the lower bits of the collapse count as output while the LSB is dropped to eliminate sensitivity to mismatch in the counter sampling flip-flop. In our work we consider $COUNT[6:4]$ as the TRNG random output bits.

\section{Hardware Trojan RO-based TRNG}
\label{sec:Trojan}
Our goal is to maliciously manipulate the TRNG design to produce predictable outputs at a particular high environmental temperature. The conditions that cause a transition from correct behavior to Trojanized behavior should be available and known only to the Trojan attacker. In order to trigger the Trojan, the attacker must apply the specific temperature which could for example be beyond the maximum operating temperature of the device.

To realize such a scenario – inspired from the stealthy parametric Trojan introduced in \cite{DBLP:conf/ches/GhandaliBHP16} – we intentionally lengthen a certain path of a combinatorial circuit. This is done in such a way that by increasing the device’s temperature, a signal on this path propagates slower than in normal operation. In the 3-edge RO construction, we achieve our goal of compromising the entropy by delaying one of the three edges of the 3-edge RO, which causes the RO to collapse in a few cycles with negligible variation. This rapid collapse behavior is not useful for generating random bits as it does not provide enough entropy.

Our technique for causing the delay change is based on manipulating one of the NAND gates that injects an edge to the RO circuit in such a way that its propagation delay is increased with temperature. The NAND gate must be very carefully altered in such a way that its propagation delay becomes more sensitive to the temperature variation than the other gates of the 3-edge RO. Note that the functionality of the design is unaltered during the normal environmental temperature.

   
In this work, we focus on manipulation of threshold voltage and show how this can be used by an attacker to trigger the Trojan at a specific operating temperature. To make the propagation delay of the target NAND gate in the 3-edge RO more sensitive to the temperature increases, we manipulate the threshold voltages of its transistors and use a combination of high $ V_{TH} $ and low $ V_{TH} $ transistors for its implementation.

\subsection{Injecting temperature-triggered Trojan into RO-based TRNG}

\begin{figure}[htbp]%
	
	\centering
	\includegraphics[width=1.0\columnwidth]{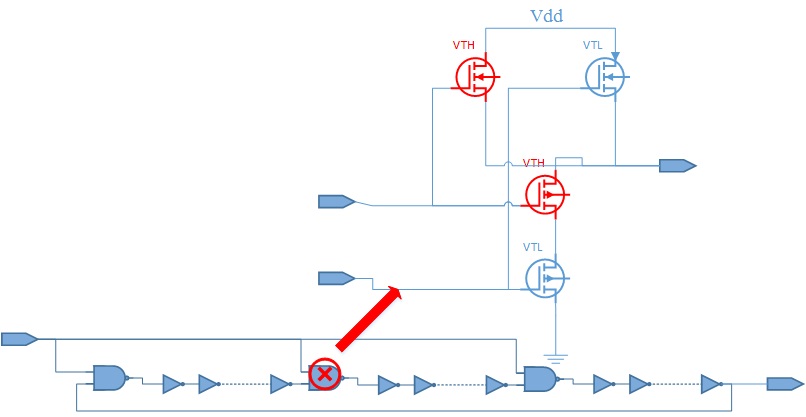}%
	\caption{Threshold voltage manipulation of the 3-edge RO}%
	\label{fig:3eRO_Trojan}%
	\vspace{-0.2 in}
\end{figure}

The time to collapse is used as the entropy source for random number generation, and delaying the start of any edges will cause the output to be not random. We focus on a single NAND gate B shown in Figure ~\ref{fig:3eRO_Trojan}. We realize the Trojan functionality by increasing the delay sensitivity of the NAND gate B to temperature increases without modifying the logic circuit. 
Edge B is injected to the RO with delay when temperature increases, so the neighboring edges of the edge B can reach it sooner than in the unmodified circuit. As a result of this, we will have a small time to collapse and hence reduced entropy. 


We show in Figure~\ref{fig:3eRO_Trojan} how we modify the transistor-level implementation of the targeted NAND gate B to make it more sensitive to the temperature. We use high threshold voltages (high-$V_{TH}$) for the NMOS and PMOS transistors connected to the start input, whose threshold voltages are increased from their standard values. As a result of this, both modified transistors will be slow to propagate the transitions on the start input to the output of the NAND gate when the temperature increases. Furthermore, to these two transistors more sensitive to temperature than the other transistors, we use low threshold voltages (low-$V_{TH}$) for the rest of the transistors in the circuit so that their delay will not increased as much as these two targeted transistors. Note that the amount of delay added to the targeted NAND gate by the threshold voltage manipulation is small in the regular environmental temperature and does not affect the behavior of the 3-edge RO, so the malicious modification is extremely difficult to detect.
\\[-5.0ex]

\begin{figure}
\centering
\subfloat{
\includegraphics[width=0.4\columnwidth]{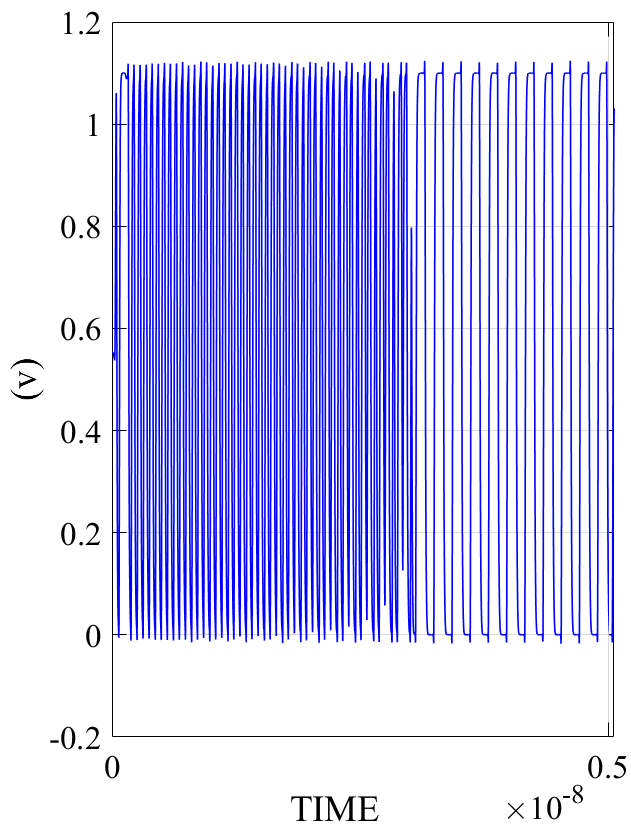}
\label{(a)}}
\subfloat{
\includegraphics[width=0.4\columnwidth]{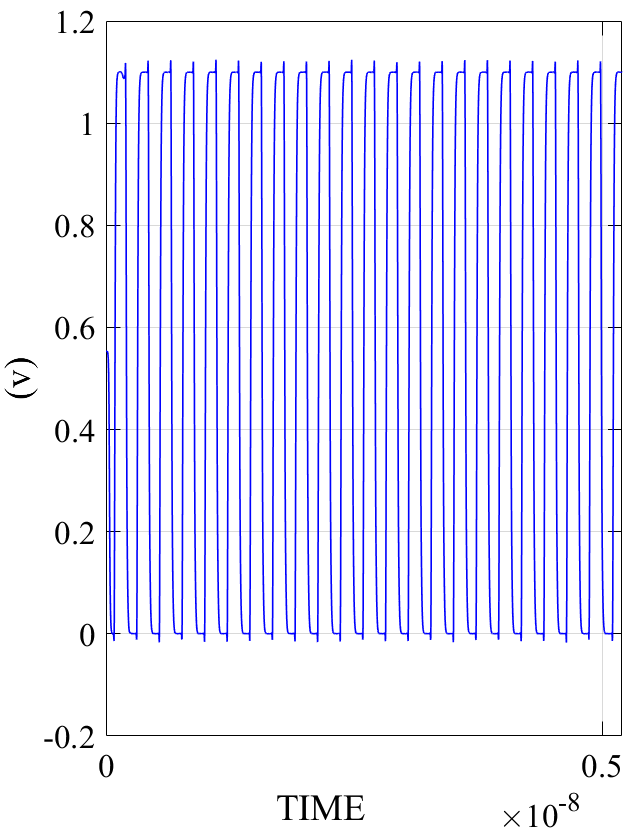}
\label{(b)}}  
\caption{Output waveform of the Trojan infected 3-edge RO at (left) 25°C where there is a large time to collapse, and (right) 120°C where there is a very small time to collapse}%
\label{fig:temperature_Trojan}%
\end{figure}

 As an example, we simulate the maliciously manipulated 3-edge RO design in two different environmental temperatures; 25°C (as a normal environmental temperature), and 120°C (as an increased environmental temperature). The Trojanized circuit behaves similar at 25°C to the unmodified 3-edge RO and there is a large collapse time (Figure~\ref{fig:temperature_Trojan}(a)) which can be used as a source of entropy for random number generation. At 120°C, the behavior of the Trojanized circuit is changed and it collapses in a few cycles (Figure~\ref{fig:temperature_Trojan}(b)). The immediate collapse occurs because the edge at NAND gate B in the manipulated 3-edge RO is not injected into the ring simultaneously with the two other edges injected at A and C. The immediate collapse behavior is not useful for extracting random bits and does not provide enough entropy. This is how the proposed temperature-triggered hardware Trojan removes the source of randomness from the 3-edge RO when the temperature rises.

\section{How to Predict the Output of the Trojan TRNG}
\label{sec:prediction}
In this section, we describe how, in principle, an attack on the Trojan infected random number generator can be executed. When an attacker wants to attack the TRNG, she may choose the environment temperature and the input master clock (MCLK) of the TRNG at her will. But even when attacker knows the operating conditions of the TRNG, its output bit-stream cannot be predicted perfectly, because of existing jitter in the TRNG, which follows independent normal distribution ($ N(0 , \sigma^{2}_{jitter}) $). 
We elaborate a stochastic model for the attacker’s knowledge to predict the output of the Trojan infected TRNG with a Markov chain model to describe the probability of occurrence for different output sequences of the Trojan infected TRNG.

\begin{figure}[htbp]%
	\centering
	\includegraphics[width=0.9\columnwidth]{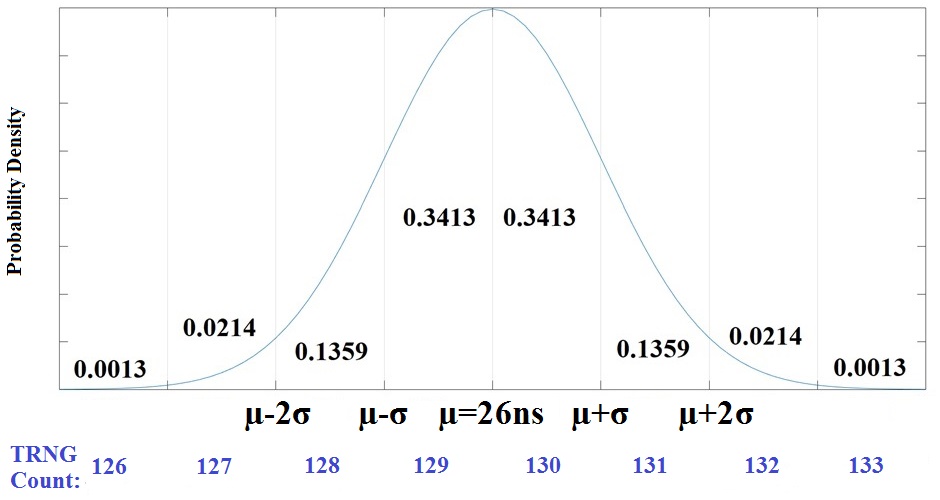}%
	\caption{Jitter effect on the Trojan infected TRNG counter values}%
	\label{fig:jitter_randbit2}%
\end{figure}

A Markov chain is a stochastic model which describes a sequence of possible events in which the probability of each event depends only on the state in the previous event~\cite{gagniuc2017markov}. Assume we have a process with a set of states S = {$ s_{1},s_{2},...,s_{r} $}. The process starts in one of these states (initial state) and moves from one state to another. If the process is in state $ s_{i} $, then it moves to state $ s_{j} $ with a transition probability $ p_{ij} $ at the next step , which is independent of states the chain was in before. The transition probabilities of all possible transitions in a Markov model can be shown by a matrix called transition matrix. Let P be the transition matrix of a Markov chain. The ijth entry $ p^{(n)}_{ij} $ of the matrix $ P^{n} $ gives the probability that the Markov chain, starting in state $ s_{i} $, will be in state $ s_{j} $ after n steps~\cite{grinstead2012introduction}.
Our Trojan removes the entropy source of the manipulated TRNG when temperature increases so that it behaves as a non-random and predictable counter when temperature rises. For example  the TRNG counter value increments by approximately 130 in each clock cycle of 26ns.
But there is variation in the amount of count due to the jitter which follows a normal distribution. For example, assume variance of jitter is  $\sigma$=100, Figure~\ref{fig:jitter_randbit2} shows the normal distribution of the jitter and the values that the Trojan infected TRNG counts corresponding to the jitter amount. 

Transition matrix of the Trojan infected TRNG for seven lower output bits  is shown by Equation~\ref{eq:TM} where $ p_{ij} $ is the transition probability that a TRNG output value which is currently $ i $ will move to value $ j $ at the next step. For example, $ p_{01} =0.341 $ is the probability of TRNG output transition from $ 0000000 $ to $ 0000001 $. If the current output value of the TRNG is $ 0000000 $, in order to have the value $ 0000001 $ as the next output, the TRNG must increment its current value by 129 in the next clock cycle, which happens with probability of 0.341 based on Figure~\ref{fig:jitter_randbit2}. As another example, consider $ p_{10} =0.021 $ which is the probability of TRNG output transition from $ 0000001 $, to $ 0000000 $. If the current output value of the TRNG is $ 0000001 $, in order to have value $ 0000000 $ as the next value of the TRNG, the TRNG must increment its count by 127, which happens with probability 0.021 as shown in Figure~\ref{fig:jitter_randbit2}.

\begin{equation}
\footnotesize
\label{eq:TM}
P =\bordermatrix{
		& 0000000      & 0000001   & \centering\dots   & 1111110      & 1111111      \cr
		0000000 & 0.136 & 0.341 & \dots & 0.001 & 0.021 \cr
		0000001 &	0.021 & 0.136 & \dots & 0.000 & 0.001 \cr
		\vdots &  \vdots & \vdots & \ddots & \vdots & \vdots \cr
		1111111 &	0.341 & 0.341 & \dots & 0.021 & 0.136                  
}_{128*128}
\normalsize
\end{equation}

 

The powers of the transition matrix of the Trojan infected TRNG give the attacker interesting information about the process as it evolves. She shall be particularly interested in the state of the chain after a large number of steps. For example, consider a scenario in which the TRNG output is used to produce a 15-bit secret key for a crypto system. Guessing this 15-bit secret key with certainty through brute force requires trying $ 2^{15} $ possible values for the key. An attacker that knows the properties of the output pattern of the Trojan infected TRNG, which are represented by the transition matrix and power matrices of the Trojan infected TRNG, can have an enhanced ability to predict output sequences.

The attacker, for guessing the 15-bit key generated by the Trojan infected TRNG, needs to predict 5 consecutive times the TRNG output ($ COUNT[6:4] $). The power matrix $ P^{4} $ gives the attacker the transition probabilities 5 steps from the current state of the TRNG output. However, the internal states between current state ($ P^{1} = P $) and the fifth state ($ P^{4} $) of the TRNG output are also important for the attacker. Assume the attacker wants to find the probability with which TRNG generates sequence $ 000,000,000,000,000 $. $ P^{4} $ gives the probabilities with which TRNG generates output value = 000 at step 5 when its output value is 000 at step 1, independent of the output values in steps 2, 3, and 4. The attacker wants to know the probability that the intermediate output values (steps 2, 3, and 4) are 000 too. To solve this problem, we modify the transition matrix $ P $ before computing $ P^{4} $ in order to avoid counting sequences that contain unwanted intermediate states. Equation~\ref{eq:modifTM} shows the modified $ P $ for sequence $ 000,000,000,000,000 $ in which we set to zero the probabilities of all unwanted transitions that are incompatible with the desired sequence. For example, transition from state 0000001 to state 1111111 corresponds to $ COUNT[6:4] = 000 $ being followed by $ COUNT[6:4] = 111 $ which is incompatible with the target sequence, so we set the transition probability to zero so that it won't be counted. As can be seen in this figure, only a block of size $ 16 \times 16 $ remains as non-zero; this $ 16 \times 16 $ block denotes the probabilities of all possible transitions from states 000xxxx to states 000xxxx where $ x \in \{0, 1\} $. After obtaining the modified transition matrix $P^\prime$, we compute $ P^{\prime^4} $  which includes the probabilities of four transitions from the current state.    
\vspace{-0.2 in}

\begin{equation*}
\label{eq:modifTM}
\resizebox{.9\hsize}{!}{$P^\prime =
\begin{blockarray}{cccccccc}
	& 0000000& 0000001& \centering\dots& 0001111& 0010000 & \centering\dots & 1111111\\
	\begin{block}{c(ccccccc)}
	0000000 & 0.136 & 0.341 &  \dots & 0.000 & 0.000 &  \dots & 0.000 \\
	0000001 & 0.021 & 0.136  &  \dots & 0.000 & 0.000 &  \dots & 0.000 \\
	\vdots & \vdots& \vdots & \ddots & \vdots & 0.000 &  \ddots & 0.000 \\
	0001111 & 0.000 & 0.000 & \dots & 0.136 & 0.000 & \dots & 0.000 \\
	0010000 & 0.000 & 0.000 & 0.000 & 0.000 & 0.000 & \dots & 0.000 \\
	\vdots &  \vdots & \vdots &  \vdots & \vdots & \vdots & \ddots & \vdots  \\
	1111111 & 0.000 & 0.000 & \dots & 0.000 & 0.000 & \dots & 0.000 \\
	\end{block}
\end{blockarray}_{128*128}$}
\end{equation*}


Consider $ u $ as the probability vector which represents the initial state of a Markov chain, then the $ i $th component of $ u $ represents the probability that the chain starts in state $ s_{i} $. For our Trojan infected TRNG we assume all initial states are equally likely to occur. The following vector represents the initial state of our manipulated TRNG in which the probability that the chain starts in any state is $ \dfrac{1}{128} $.

\begin{equation}
	\label{eq:initialstate}
	u = [\dfrac{1}{128} \dfrac{1}{128} \dfrac{1}{128} ... \dfrac{1}{128} \dfrac{1}{128}]_{1 \times 128}
\end{equation}

The probability that the chain is in state $ s_{i} $ after $ n $ steps is the $ i $th entry in the following vector:

\begin{equation}
	\label{eq:nthstate}
	u^{(n)} = uP^{n}
\end{equation}

To obtain the probability of the sequence $ 000,000,000,000,000 $ we set n = 4 in the Equation~\ref{eq:nthstate} and then add all non-zero probabilities as shown in Equations~\ref{eq:all0} and ~\ref{eq:all0P}. The obtained value is almost equal to the measured value in our experiment.
  
\begin{equation}
\label{eq:all0}
u^{(4)} = u{P^\prime}^{4} = [\dfrac{1}{128} \dfrac{1}{128} \dfrac{1}{128} ... \dfrac{1}{128} \dfrac{1}{128}]{P^\prime}^{4}
\end{equation}  
\begin{equation}
	\label{eq:all0P}
	P(000,000,000,000,000) = \sum^{i=128}_{i=0} u{P^\prime}^{4}[i] = 0.0764
\end{equation} 

An attacker can use this method to obtain the most likely patterns for an n-bit key. Table~\ref{tab:15bitpattern} lists the eight most likely patterns of a 15-bit key and their probabilities. The attacker can guess the 15-bit key with the probability of 0.61 by trying these eight patterns.   

\begin{table}[htbp]
	\vspace{-0.1 in}
	\centering
	\caption{Most likely 15-bit patterns}
	\label{tab:15bitpattern}
	\begin{tabular}{ c c }
		\specialrule{.3em}{.2em}{.2em}
		15-bit Pattern  & Probability \\ \specialrule{.3em}{.2em}{.2em} 
		000000000000000 & 0.0764      \\ \hline
		001001001001001 & 0.0764      \\ \hline
		010010010010010 & 0.0764      \\ \hline
		011011011011011 & 0.0764      \\ \hline
		100100100100100 & 0.0764      \\ \hline
		101101101101101 & 0.0764      \\ \hline
		110110110110110 & 0.0764      \\ \hline
		111111111111111 & 0.0764      \\ \specialrule{.3em}{.2em}{.2em}
	\end{tabular}
\vspace{-0.1 in}
\end{table}

\section{Practical Results}
\label{sec:result}
45nm Nangate Open Cell Library is used for our implementation of the Trojan free and Trojan infected TRNGs.

 The randomness of the Trojan free TRNG and the Trojan infected TRNG are evaluated by the NIST statistical test suite \cite{bassham2010statistical}. The Trojan free TRNG is robust and passes all NIST tests across all temperatures (25°C, 60°C, 120°C) as shown in Table~\ref{tab:NISTres}.
 The NIST test suite results of the Trojan infected TRNG are also shown in this table for different temperatures (25°C, 60°C, 120°C). The Trojan infected TRNG passes the tests at the normal environmental temperatures (25°C, 60°C), but at the trigger temperature of 120°C does not pass the tests.

 \begin{table}[htbp]
 	\centering
 	
 	\caption{NIST test suite results for Trojan free and Trojan infected TRNG }\label{tab:NISTres}
 	\begin{tabular}{|c|p{0.2in}|p{0.2in}|p{0.2in}|p{0.2in}|p{0.2in}|p{0.2in}|}
 		\hline
 		\multirow{2}{*}{NIST} & \multicolumn{3}{c|}{Trojan free design} & \multicolumn{3}{c|}{Trojan infected design} \\ \cline{2-7} 
 		& 25°C & 60°C & 120°C & 25°C & 60°C & 120°C \\ \hline
 		Frequency & pass & pass & pass & pass & pass & pass \\ \hline
 		Block frequency & pass & pass & pass & pass & pass & fail \\ \hline
 		Cumulative sums (1) & pass & pass & pass & pass & pass & fail \\ \hline
 		Cumulative sums (2) & pass & pass & pass & pass & pass & pass \\ \hline
 		Longest runs & pass & pass & pass & pass & pass & pass \\ \hline
 		FFT & pass & pass & pass & pass & pass & fail \\ \hline
 		Approximate entropy & pass & pass & pass & pass & pass & fail \\ \hline
 	\end{tabular}
 \end{table}

\begin{figure}[h]%
	\centering	
	\includegraphics[width=1.0\columnwidth]{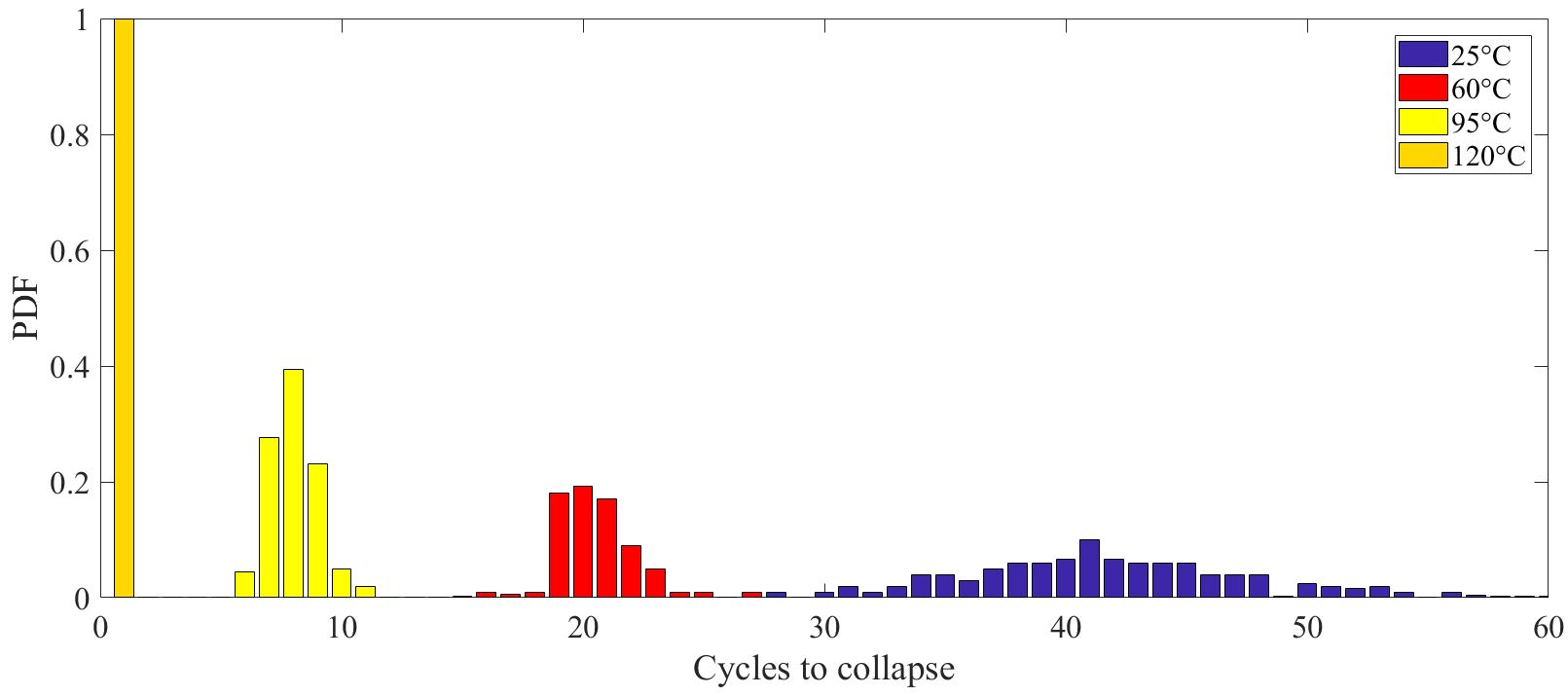}%
	\caption{Distribution of 3-edge RO cycles to collapse at different environmental temperatures}%
	\vspace{-0.2 in}
	\label{fig:Cycle_to_collapse_PDF_temp}%
\end{figure}

The measured distribution of number of cycles to collapse of the Trojan infected 3-edge RO at different environmental temperatures are shown in Figure~\ref{fig:Cycle_to_collapse_PDF_temp} which follows inverse Gaussian distribution. Increasing the temperature causes the mean and variance of the number of cycles to collapse to decrease. At 120°C the mean value becomes 0 with negligible variance, meaning that the Trojan infected TRNG collapses within the first few cycles and therefore does not provide enough entropy.  

\begin{figure}[htbp]
	\centering
	\subfloat[Trojan free TRNG]{\includegraphics[clip,width=0.8\linewidth]{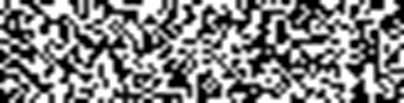}\label{fig:f1}}
	\vspace{-0.1 in}
	\vfill
	\subfloat[Trojan infected TRNG]{\includegraphics[clip,width=0.8\linewidth]{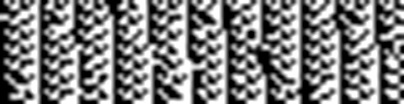}\label{fig:f2}}	
	\caption{Output patterns of (a) the Trojan free TRNG, and (b) the Trojan infected TRNG, raster scanning left-to-right then top-to-bottom.}%
	\label{fig:TRNGbitstream}%
\end{figure}

\begin{figure}[htbp]%
	\centering
	\begin{tabular}{@{}c@{}}
		\vspace{-0.1 in}
		\includegraphics[width=0.85\linewidth]{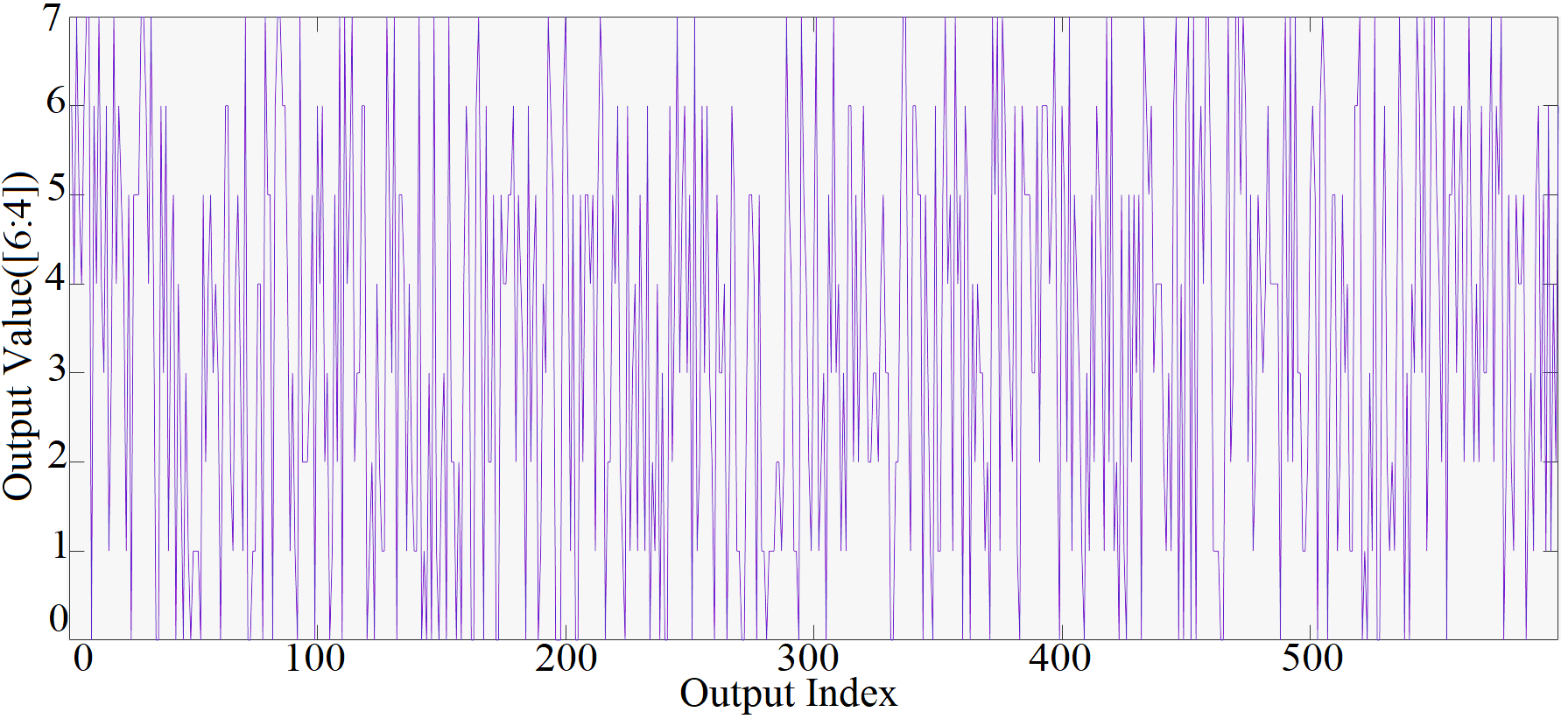} \\[\abovecaptionskip]			
		\small (a)
		
	\end{tabular}	
	\begin{tabular}{@{}c@{}}
		\vspace{-0.1 in}
		\includegraphics[width=0.85\linewidth]{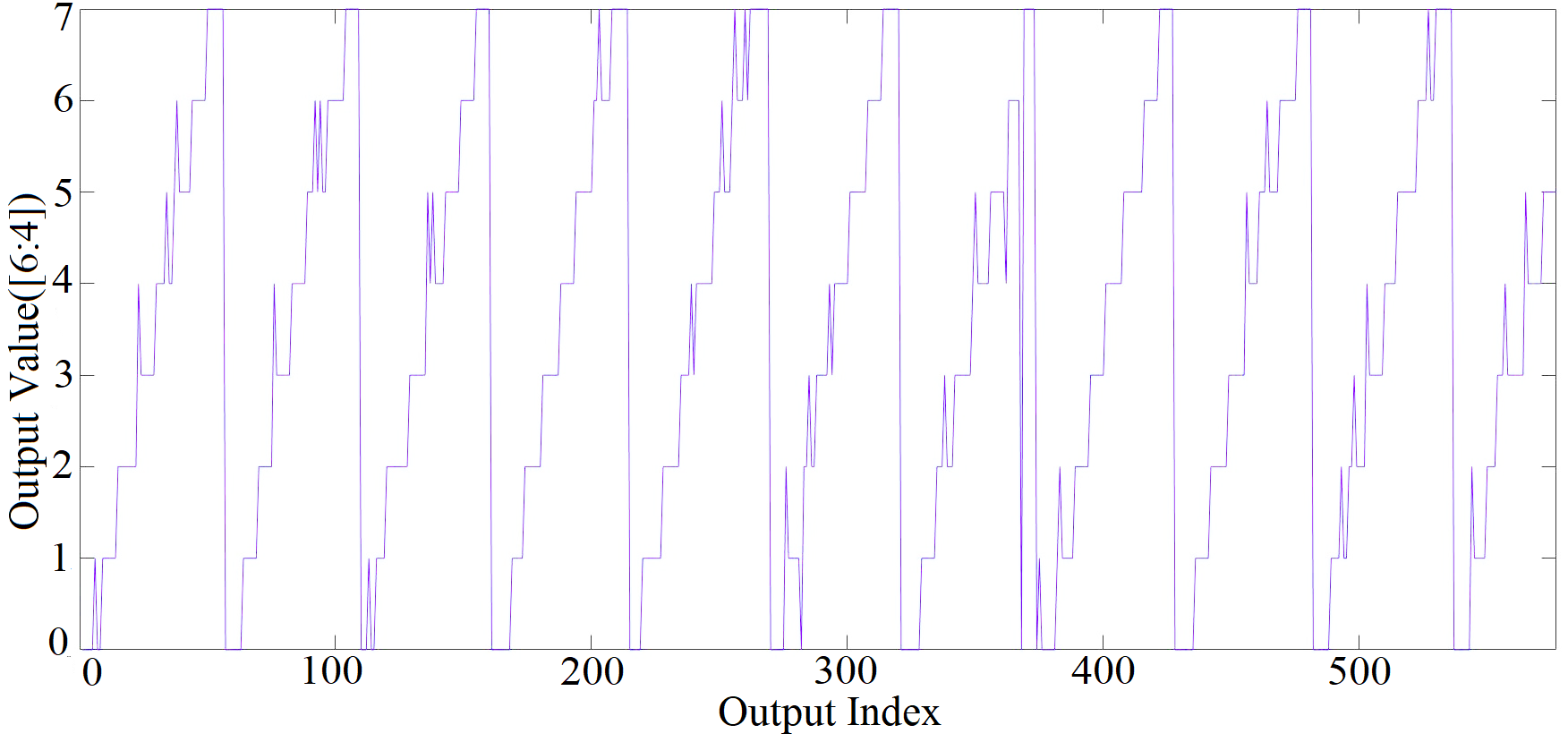} \\[\abovecaptionskip]		
		\small (b)
		\vspace{-0.1 in}
	\end{tabular}
	\caption{Output values of (a) Trojan free TRNG and (b) Trojan infected TRNG at 120°C.}%
	\label{fig:TRNGoutvalue}%
\end{figure}

Figure~\ref{fig:TRNGbitstream} illustrates the Trojan free TRNG bitstream and also the Trojan infected TRNG bitstream, raster scanning top-to-bottom then left-to-right. The outputs of the Trojan-free TRNG do not have any apparent pattern (Figure~\ref{fig:TRNGbitstream}(a)), while the outputs of the Trojan infected TRNG at the trigger temperature are clearly periodic and non-random (Figure~\ref{fig:TRNGbitstream}(b)). As another view of the same data, the output values of the Trojan-free TRNG for 600 samples (1800 bits) are shown in Figure~\ref{fig:TRNGoutvalue}(a), and the output values of the Trojan infected TRNG are shown in Figure~\ref{fig:TRNGoutvalue}(b). The Trojan infected TRNG produces output patterns that are largely periodic but have some noise.

 \begin{table}[htb]
 	\caption{Attack complexities of different key sizes}
 	\label{tab:attackcomp}
 	\centering
 	\begin{tabular}{c|c|c}
 		\specialrule{.3em}{.2em}{.2em}
 		& Attack Complexity & Probability of success \\ \specialrule{.3em}{.2em}{.2em}
 		\multirow{2}{*}{64-bit} & $2^{15}$ & 0.8928 \\ \cline{2-3}\\[-1em] 
 		& $2^{18}$ & 0.996 \\ \hline\\[-1em]
 		\multirow{2}{*}{128-bit} & $2^{25}$ & 0.85 \\ \cline{2-3} \\[-1em]
 		& $2^{29}$ & 0.98 \\ \hline\\[-1em]
 		\multirow{2}{*}{256-bit} & $2^{55}$ & 0.964 \\ \cline{2-3} \\[-1em]
 		& $2^{58}$ & 0.9872 \\ \specialrule{.3em}{.2em}{.2em}
 	\end{tabular}
 	\\[-5.0ex]
 \end{table}
Table~\ref{tab:attackcomp} reports the attack complexities of 64-bit, 128-bit, and 256-bit keys for two probability values. There is a trade off between the attack complexity and the probability of success, but as can be seen in the table, for lower attack complexities of each key size, the probability of success is still high and acceptable.

\section{Conclusions}
\label{sec:conclude}
\vspace{-0.1 in}
We show how a parametric hardware Trojan with very low overhead can be inserted into RO-based TRNG designs. This kind of parametric Trojan is very hard to be detected, because in general it does not require the addition or removal of any logic into or from the target design. Thus, even in a white-box scenario the Trojan remains stealthy and is unlikely to be detected by an evaluation lab.
The underlying concept is based on removing source of entropy of the TRNG when Trojan is triggered in high temperature, while the malicious TRNG works correctly and generate random outputs in normal conditions. To inject the Trojan, we lengthen the certain path of a combinatorial logic realizing RO in such a way that by increasing temperature the source of entropy is not enough anymore. We elaborate a stochastic model based on Markov Chain for the attacker’s knowledge to predict the output of the Trojan infected TRNG. This parametric Trojan allows us to significantly lower the security level even of highly protected crypto-core implementations that are connected to the TRNG.







\bibliographystyle{splncs03}
\bibliography{bib/abbrev3,bib/crypto_custom,bib/eprint,bib/SCA,bib/HT,bib/other,bib/trng,bib/cryptanalysis,bib/trojan}

\begin{thebibliography}{10}
\providecommand{\url}[1]{\texttt{#1}}
\providecommand{\urlprefix}{URL }

\bibitem{bassham2010statistical}
Bassham, L.E., Rukhin, A.L., Soto, J., Nechvatal, J.R., Smid, M.E., Leigh,
  S.D., Levenson, M., Vangel, M., Heckert, N.A., Banks, D.L.: A statistical
  test suite for random and pseudorandom number generators for cryptographic
  applications. Tech. rep. (2010)

\bibitem{conf/cosade/BayonBAFPRM12}
Bayon, P., Bossuet, L., Aubert, A., Fischer, V., Poucheret, F., Robisson, B.,
  Maurine, P.: In: COSADE. Lecture Notes in Computer Science, vol. 7275, pp.
  151--166. Springer (2012)

\bibitem{DBLP:conf/ches/BeckerRPB13}
Becker, G.T., Regazzoni, F., Paar, C., Burleson, W.P.: {Stealthy Dopant-Level
  Hardware Trojans}. In: CHES 2013. Lecture Notes in Computer Science, vol.
  8086, pp. 197--214. Springer (2013)

\bibitem{DBLP:conf/crypto/BihamCS08}
Biham, E., Carmeli, Y., Shamir, A.: {Bug Attacks}. In: CRYPTO 2008. Lecture
  Notes in Computer Science, vol. 5157, pp. 221--240. Springer (2008)

\bibitem{ChakrabortyNB09}
Chakraborty, R.S., Narasimhan, S., Bhunia, S.: {Hardware Trojan: Threats and
  emerging solutions}. In: HLDVT 2009. pp. 166--171. {IEEE} Computer Society
  (2009)

\bibitem{DBLP:conf/ches/CherkaouiFFA13}
Cherkaoui, A., Fischer, V., Fesquet, L., Aubert, A.: A very high speed true
  random number generator with entropy assessment. In: Cryptographic Hardware
  and Embedded Systems - {CHES} 2013 - 15th International Workshop, Santa
  Barbara, CA, USA, August 20-23, 2013. Proceedings. pp. 179--196 (2013)

\bibitem{EnderG0P17}
Ender, M., Ghandali, S., Moradi, A., Paar, C.: The first thorough side-channel
  hardware trojan. In: 23rd International Conference on the Theory and
  Applications of Cryptology and Information Security(ASIACRYPT). pp. 755--780
  (2017)

\bibitem{gagniuc2017markov}
Gagniuc, P.A.: Markov chains: from theory to implementation and
  experimentation. John Wiley \& Sons (2017)

\bibitem{DBLP:conf/ches/GhandaliBHP16}
Ghandali, S., Becker, G.T., Holcomb, D., Paar, C.: {A Design Methodology for
  Stealthy Parametric Trojans and Its Application to Bug Attacks}. In: CHES
  2016. Lecture Notes in Computer Science, vol. 9813, pp. 625--647. Springer
  (2016)

\bibitem{grinstead2012introduction}
Grinstead, C.M., Snell, J.L.: Introduction to probability. American
  Mathematical Soc. (2012)

\bibitem{DBLP:conf/host/JinM08}
Jin, Y., Makris, Y.: {Hardware Trojan Detection Using Path Delay Fingerprint}.
  In: HOST 2008. pp. 51--57. {IEEE} Computer Society (2008)

\bibitem{DBLP:conf/fdtc/KumarJBP14}
Kumar, R., Jovanovic, P., Burleson, W.P., Polian, I.: {Parametric Trojans for
  Fault-Injection Attacks on Cryptographic Hardware}. In: FDTC 2014. pp.
  18--28. {IEEE} Computer Society (2014)

\bibitem{DBLP:conf/iccad/LinBP09}
Lin, L., Burleson, W., Paar, C.: {MOLES: Malicious off-chip leakage enabled by
  side-channels}. In: ICCAD 2009. pp. 117--122. {ACM} (2009)

\bibitem{liu2011true}
Liu, N., Pinckney, N., Hanson, S., Sylvester, D., Blaauw, D.: A true random
  number generator using time-dependent dielectric breakdown. In: 2011
  Symposium on VLSI Circuits-Digest of Technical Papers. pp. 216--217. IEEE
  (2011)

\bibitem{DBLP:conf/ches/MarkettosM09}
Markettos, A.T., Moore, S.W.: The frequency injection attack on
  ring-oscillator-based true random number generators. In: Cryptographic
  Hardware and Embedded Systems - {CHES} 2009, 11th International Workshop,
  Lausanne, Switzerland, September 6-9, 2009, Proceedings. pp. 317--331 (2009)

\bibitem{martin2015fault}
Martin, H., Korak, T., San~Milln, E., Hutter, M.: Fault attacks on strngs:
  Impact of glitches, temperature, and underpowering on randomness. IEEE
  transactions on information forensics and security  10(2),  266--277 (2015)

\bibitem{DBLP:conf/ahs/ShiyanovskiiWRPWC10}
Shiyanovskii, Y., Wolff, F.G., Rajendran, A., Papachristou, C.A., Weyer, D.J.,
  Clay, W.: {Process reliability based trojans through NBTI and HCI effects}.
  In: Adaptive Hardware and Systems {AHS} 2010. pp. 215--222. {IEEE} (2010)

\bibitem{DBLP:conf/cicc/TangKLPK14}
Tang, Q., Kim, B., Lao, Y., Parhi, K.K., Kim, C.H.: True random number
  generator circuits based on single- and multi-phase beat frequency detection.
  In: Proceedings of the {IEEE} 2014 Custom Integrated Circuits Conference,
  {CICC} 2014, San Jose, CA, USA, September 15-17, 2014. pp. 1--4 (2014)

\bibitem{DBLP:conf/dft/WangSTP08}
Wang, X., Salmani, H., Tehranipoor, M., Plusquellic, J.F.: {Hardware Trojan
  Detection and Isolation Using Current Integration and Localized Current
  Analysis}. In: DFT 2008. pp. 87--95. {IEEE} Computer Society (2008)

\bibitem{DBLP:journals/ijrc/WoldT09}
Wold, K., Tan, C.H.: Analysis and enhancement of random number generator in
  {FPGA} based on oscillator rings. Int. J. Reconfig. Comp.  2009,
  501672:1--501672:8 (2009), \url{https://doi.org/10.1155/2009/501672}

\bibitem{yang201416}
Yang, K., Fick, D., Henry, M.B., Lee, Y., Blaauw, D., Sylvester, D.: 16.3 a
  23mb/s 23pj/b fully synthesized true-random-number generator in 28nm and 65nm
  cmos. In: Solid-State Circuits Conference Digest of Technical Papers (ISSCC),
  2014 IEEE International. pp. 280--281. IEEE (2014)

\end{thebibliography}

\end{document}